\documentclass[11pt]{article}

\usepackage[round]{natbib}
\usepackage{a4}
\usepackage{color}
\usepackage{hyperref}
\usepackage{graphicx}
\usepackage{caption}
\usepackage{lscape}
\extrafloats{100}
\usepackage{tablefootnote}

\begin{document}

\begin{center}
\begin{tabular}{lcccc}
\Large {\bf GEOS}  & \hspace{1.8cm} & GEOS CIRCULAR ON RR TYPE VARIABLE   & \hspace{1.2cm}& GEOS \\
\Large {\bf RR 58} &                & Oct. 2018                           &             & 23 Parc de Levesville \\
                   &                &                          \multicolumn{3}{r}{ F-28300 BAILLEAU L'EV\^EQUE }\\
\end{tabular}
\end{center}
\vspace{0.5cm}

\begin{center}
\Large
{\bf CCD and DSLR maxima of RR Lyrae stars in 2016 and 2017}
\vspace{0.5cm}

\Large
{\it Joint publication: VOSJ Variable Star Bulletin No.65} \\
\vspace{0.5cm}
\end{center}

\normalsize
Kenji Hirosawa$^{1}$ and J.F. Le Borgne$^{2}$\\
\normalsize
$^{1}$ Variable Star Observer League in Japan,\\
$^{2}$ Groupe Europ\'een d’Observations Stellaires\\

\vspace{0.6cm}

We present here maximum timings of RR Lyrae stars observed in 2016 and 2017
by VSOLJ member Kenji Hirosawa.
The calculation of the time of maximum from the measurements provided by the
observer have been made in the frame of the GEOS RR Lyr survey.
GEOS (European group for stellar observation, http://geos.upv.es/) is an association
of amateur and
professional astronomers, created in 1974, the aim of which is the promotion of
research in astrophysics by amateur astronomers. A natural domain of research
of the group is the study of variable stars. In this context, it appeared that
the observation of the stars of RR Lyrae type is relatively neglected, in particular
on long time scales. This is why GEOS decided about one decade ago to start a "RR Lyr
Survey" (GRRS). It includes 3 main parts: \\
- A routine survey of bright RR Lyr stars, aiming in measuring times of maximum of
RRab stars brighter than magnitude about 13 in order to monitor the variation of
period on decade time scale. The time sampling is of the order of 10 maximums per year for
each star. The observations are mainly done by professional robotic telescopes
(25 cm TAROT telescopes, \cite{Klotz}), but also by amateur astronomers performing
telescopes of 10 to 25cm diameter equipped with CCD or DSLR cameras
\citep{Le Borgne 2007}. \\
- Prospecting survey of under-studied RR Lyraes of magnitudes 13 to 15 using
telescopes of 20 to 60 cm equipped with CCD cameras. For this part the aim is to
check variation type, period and possible Blazhko effect. \\
- The follow-up of the RR Lyr it self in order to monitor the evolution of its
Blazhko effect. Preston et al. \citep{Preston} have shown that in years 1960's the
Blazhko effect of RR Lyr have stopped and restarted again afterwards. This phenomenon
was not studied in details because it requests continuous monitoring of the star.
GRRS observers use dedicated small instruments (5cm photographic lens with CCD
or DSLR) to follow RR Lyr since 2008. The time sampling of times of maximum
measurements have allow to observe the disappearance of RR Lyr Blazhko effect in
2014 \citep{Le Borgne 2014} \citep{Poretti}. It restarted afterwards but with very
small amplitude, and in 2018, the amplitude of the Blazhko effect is not still
at the value it had at the beginning of the survey. \\
GRRS contributes to the "GEOS RR Lyr database" (http://rr-lyr.irap.omp.eu/dbrr/)

\newpage

The O-C are computed using the elements from the General Catalogue of Variable
Stars \citep{GCVS} by default. For the stars which have no elements in GCVS, the following
table give the list of elements used. \\

\vspace{1.cm}

\begin{center}
\small
\begin{tabular}{llll}\hline
Star & Origin & Period & reference\\
\hline
V1962 Cyg           & 2455729.470  & 0.508337   &  this paper  \\
BN Eri              & 2457358.054  & 0.4876848  &  this paper  \\
V552 Her            & 2457153.220  & 0.3785194  &  this paper  \\
IY Peg              & 2450698.0011 & 0.54572621 &  \cite{Vandenbroere} \\
EX UMa              & 2455228.4423 & 0.5428328  &  this paper  \\
ASAS J020058+1332.8 & 2457697.191  & 0.616166   &  this paper  \\
NSV 1443            & 2457641.2922 & 0.6088106  &  this paper  \\
NSV 3833            & 2457697.304  & 0.48724    &  this paper  \\
\hline
\end{tabular}
\normalsize
\end{center}

\vspace{1.cm}

The following tables contain the list of maximums observed by Kenji Hirosawa in
2016 and 2017.

\begin{center}
\small
\begin{tabular}{lrrrrlrll}\hline
star & max.HJD & O-C & err & E & band & n obs. & code & instrument\\
\hline
SW And              & 2457974.1717 & -0.0034 &  0.0029 &   9584 & cG & 126& Hsk& CANON X7 \\
XX And              & 2458090.9066 & -0.0005 &  0.0019 &   6127 & V  & 125& Hsk& 20cmL+ST402ME \\
ZZ And              & 2457671.1584 &  0.0214 &  0.0019 &   7835 & V  & 138& Hsk& 20cmL+ST402ME \\
BK And              & 2457641.2051 & -0.0140 &  0.0011 &  16368 & V  & 107& Hsk& 20cmL+ST402ME \\
CI And              & 2457403.9242 & -0.0172 &  0.0012 &  12300 & V  & 173& Hsk& 20cmL+ST402ME \\
CI And              & 2457697.1868 & -0.0143 &  0.0019 &  12905 & V  & 123& Hsk& 13cmR+ST8XME \\
DE And              & 2457668.1732 &  0.0396 &  0.0020 &  21064 & V  &  82& Hsk& 25cmL+ST402ME \\
DR And              & 2458015.1054 & -0.1185 &  0.0018 &   9580 & V  & 132& Hsk& 25cmL+ST402ME \\
DR And              & 2458091.1306 & -0.1159 &  0.0023 &   9715 & V  & 110& Hsk& 25cmL+ST402ME \\
DY And              & 2458090.9979 & -0.0865 &  0.0022 &  12327 & V  & 113& Hsk& 25cmL+ST402ME \\
OV And              & 2458091.0880 & -0.0115 &  0.0020 &  10106 & V  & 125& Hsk& 13cmR+ST8XME \\
SW Aqr              & 2458051.9188 &  0.0356 &  0.0008 &  11224 & V  & 148& Hsk& 20cmL+ST402ME \\
SX Aqr              & 2458027.0644 & -0.0061 &  0.0020 &   7736 & V  & 144& Hsk& 20cmL+ST402ME \\
SX Aqr              & 2458064.0283 & -0.0061 &  0.0019 &   7805 & V  &  94& Hsk& 13cmR+ST8XME \\
TZ Aqr              & 2458063.9335 &  0.0393 &  0.0022 &   9001 & V  & 115& Hsk& 25cmL+ST402ME \\
AA Aqr              & 2458081.9010 & -0.0443 &  0.0019 &   7394 & V  & 109& Hsk& 20cmL+ST402ME \\
BO Aqr              & 2458057.9124 &  0.0297 &  0.0036 &   7399 & V  & 117& Hsk& 20cmL+ST402ME \\
BR Aqr              & 2457974.1659 & -0.0039 &  0.0015 &  12255 & V  & 117& Hsk& 20cmL+ST402ME \\
CP Aqr              & 2458022.0871 &  0.0054 &  0.0008 &  12791 & V  & 141& Hsk& 25cmL+ST402ME \\
FX Aqr              & 2458078.9239 & -0.0170 &  0.0036 &   8203 & V  & 115& Hsk& 25cmL+ST402ME \\
HH Aqr              & 2457609.2566 & -0.1069 &  0.0034 &   9957 & V  & 178& Hsk& 25cmL+ST402ME \\
V341 Aql            & 2458062.9419 &  0.0048 &  0.0020 &   5696 & V  & 142& Hsk& 25cmL+ST402ME \\
SY Ari              & 2458018.2760 & -0.0287 &  0.0022 &   7819 & V  & 116& Hsk& 20cmL+ST402ME \\
SY Ari              & 2458110.0800 & -0.0271 &  0.0022 &   7982 & V  & 116& Hsk& 20cmL+ST402ME \\
&  &  &  &  &   &   &  &  \\
\hline
\end{tabular}
\normalsize
\end{center}
\begin{center}
\small
\begin{tabular}{lrrrrlrll}\hline
star & max.HJD & O-C & err & E & band & n obs. & code & instrument\\
\hline
TV Ari              & 2457974.2595 &  0.0227 &  0.0028 &   7990 & V  & 125& Hsk& 25cmL+ST402ME \\
CD Ari              & 2458082.0063 &  0.0691 &  0.0030 &  15664 & V  & 114& Hsk& 25cmL+ST402ME \\
CD Ari              & 2458109.9147 &  0.0735 &  0.0045 &  15749 & V  & 135& Hsk& 13cmR+ST8XME \\
CI Ari              & 2458025.2694 & -0.0655 &  0.0011 &  11273 & V  & 135& Hsk& 20cmL+ST402ME \\
V575 Aur            & 2457668.1829 &  0.0257 &  0.0020 &   9394 & V  &  93& Hsk& 20cmL+ST402ME \\
RS Boo              & 2457418.3570 & -0.0099 &  0.0013 &  23634 & V  &  70& Hsk& 13cmR+ST8XME \\
RS Boo              & 2457867.0109 & -0.0121 &  0.0015 &  24822 & cG & 168& Hsk& CANON X4 \\
ST Boo              & 2457421.3561 &  0.0918 &  0.0014 &  14336 & V  & 141& Hsk& 20cmL+ST402ME \\
ST Boo              & 2457762.3652 &  0.0882 &  0.0023 &  14884 & V  &  93& Hsk& 20cmL+ST402ME \\
SW Boo              & 2457418.3417 &  0.0261 &  0.0010 &   7551 & V  &  84& Hsk& 20cmL+ST402ME \\
TV Boo              & 2457421.1563 &  0.0053 &  0.0014 &  19556 & V  & 206& Hsk& 20cmL+ST402ME \\
TW Boo              & 2457743.3238 & -0.0542 &  0.0015 &  11617 & V  & 117& Hsk& 25cmL+ST402ME \\
UU Boo              & 2457496.0048 &  0.0114 &  0.0010 &   6575 & V  & 138& Hsk& 25cmL+ST402ME \\
UU Boo              & 2457872.0662 &  0.0162 &  0.0015 &   7398 & V  & 160& Hsk& 13cmR+ST8XME \\
UY Boo              & 2457444.2964 & -0.1488 &  0.0022 &   5944 & V  & 118& Hsk& 20cmL+ST402ME \\
UY Boo              & 2457463.1560 & -0.1652 &  0.0015 &   5973 & V  &  84& Hsk& 13cmR+ST8XME \\
WW Boo              & 2457512.9777 & -0.0285 &  0.0019 &   7544 & V  & 101& Hsk& 25cmL+ST402ME \\
CM Boo              & 2457872.1100 & -0.0072 &  0.0025 &   6583 & V  & 162& Hsk& 25cmL+ST402ME \\
KR Boo              & 2457758.2836 &  0.0469 &  0.0015 &   9614 & V  &  83& Hsk& 20cmL+ST402ME \\
LW Boo              & 2457857.2507 &  0.0776 &  0.0037 &   6534 & V  & 116& Hsk& 20cmL+ST402ME \\
RZ Cam              & 2457434.9349 & -0.0004 &  0.0023 &   6012 & V  & 129& Hsk& 25cmL+ST402ME \\
RZ Cam              & 2457671.3138 & -0.0036 &  0.0040 &   6504 & V  & 108& Hsk& 25cmL+ST402ME \\
RZ Cam              & 2457752.9924 & -0.0017 &  0.0016 &   6674 & V  & 114& Hsk& 20cmL+ST402ME \\
RZ Cam              & 2458082.0983 & -0.0051 &  0.0015 &   7359 & V  &  99& Hsk& 20cmL+ST402ME \\
LP Cam              & 2457752.9004 &  0.1621 &  0.0032 &   6972 & V  & 120& Hsk& 20cmL+ST402ME \\
RW Cnc              & 2457750.3557 & -0.0345 &  0.0020 &  11834 & V  &  71& Hsk& 20cmL+ST402ME \\
RW Cnc              & 2457753.0935 & -0.0327 &  0.0015 &  11839 & V  & 122& Hsk& 13cmR+ST8XME \\
SS Cnc              & 2457421.0768 &  0.0145 &  0.0008 &   7969 & V  & 139& Hsk& 20cmL+ST402ME \\
AN Cnc              & 2457780.0839 &  0.0177 &  0.0027 &   7512 & V  &  94& Hsk& 20cmL+ST402ME \\
AQ Cnc              & 2457435.0496 &  0.0062 &  0.0014 &   7852 & V  & 170& Hsk& 25cmL+ST402ME \\
AS Cnc              & 2457404.0726 & -0.0121 &  0.0023 &   7715 & V  & 138& Hsk& 25cmL+ST402ME \\
AS Cnc              & 2457815.9601 & -0.0210 &  0.0014 &   8382 & V  & 103& Hsk& 25cmL+ST402ME \\
CQ Cnc              & 2457459.9624 & -0.0090 &  0.0013 &   5553 & V  & 120& Hsk& 25cmL+ST402ME \\
CQ Cnc              & 2457780.0012 & -0.0073 &  0.0022 &   6163 & V  & 100& Hsk& 25cmL+ST402ME \\
EZ Cnc              & 2457717.3087 &  0.0088 &  0.0013 &   7191 & V  & 120& Hsk& 20cmL+ST402ME \\
KV Cnc              & 2457422.0228 & -0.0384 &  0.0032 &   9284 & V  & 209& Hsk& 25cmL+ST402ME \\
W CVn               & 2457422.2180 & -0.0778 &  0.0017 &  11136 & V  & 175& Hsk& 13cmR+ST8XME \\
Z CVn               & 2457866.9877 & -0.2069 &  0.0047 &  28960 & V  & 147& Hsk& 20cmL+ST402ME \\
RR CVn              & 2457476.0663 &  0.0260 &  0.0014 &  24714 & V  & 130& Hsk& 20cmL+ST402ME \\
RR CVn              & 2457872.1186 &  0.0255 &  0.0013 &  25423 & V  & 161& Hsk& 20cmL+ST402ME \\
RU CVn              & 2457496.0620 & -0.0165 &  0.0020 &   5674 & V  &  80& Hsk& 13cmR+ST8XME \\
RX CVn              & 2457475.9710 &  0.0262 &  0.0025 &  10955 & V  & 100& Hsk& 20cmL+ST402ME \\
RZ CVn              & 2457495.9931 &  0.1960 &  0.0017 &  10906 & V  & 150& Hsk& 20cmL+ST402ME \\
SS CVn              & 2457450.1266 &  0.2321 &  0.0017 &  12783 & V  & 153& Hsk& 20cmL+ST402ME \\
SW CVn              & 2457444.3039 & -0.1220 &  0.0015 &  40503 & V  & 133& Hsk& 25cmL+ST402ME \\
SZ CVn              & 2457450.0721 &  0.0011 &  0.0022 &   3959 & V  & 157& Hsk& 25cmL+ST402ME \\
X CMi               & 2457721.2960 &  0.0199 &  0.0024 &   8907 & V  & 144& Hsk& 20cmL+ST402ME \\
AA CMi              & 2457780.0618 &  0.1040 &  0.0019 &  44515 & V  & 134& Hsk& 13cmR+ST8XME \\
AL CMi              & 2457421.0788 &  0.0045 &  0.0019 &   4640 & V  & 156& Hsk& 25cmL+ST402ME \\
AL CMi              & 2457780.0145 &  0.0066 &  0.0022 &   5292 & V  & 100& Hsk& 20cmL+ST402ME \\
&  &  &  &  &   &   &  &  \\
\hline
\end{tabular}
\normalsize
\end{center}
\begin{center}
\small
\begin{tabular}{lrrrrlrll}\hline
star & max.HJD & O-C & err & E & band & n obs. & code & instrument\\
\hline
DQ CMi              & 2457404.0815 &  0.0622 &  0.0033 &   7100 & V  & 147& Hsk& 20cmL+ST402ME \\
DQ CMi              & 2457676.2856 &  0.0694 &  0.0040 &   7542 & V  & 152& Hsk& 20cmL+ST402ME \\
HU Cas              & 2457421.9123 & -0.0277 &  0.0017 &  14379 & V  & 150& Hsk& 20cmL+ST402ME \\
HU Cas              & 2457611.2418 & -0.0315 &  0.0013 &  14839 & V  & 133& Hsk& 20cmL+ST402ME \\
HU Cas              & 2458063.9989 & -0.0281 &  0.0015 &  15939 & V  & 118& Hsk& 25cmL+ST402ME \\
IU Cas              & 2457999.2396 &  0.0448 &  0.0021 &  12880 & V  & 153& Hsk& 20cmL+ST402ME \\
RZ Cep              & 2458026.9839 & -0.0565 &  0.0050 &  49862 & cG & 182& Hsk& CANON X4 \\
RZ Cep              & 2458051.9832 & -0.0607 &  0.0023 &  49943 & cG & 181& Hsk& CANON X4 \\
RZ Cep              & 2458056.9270 & -0.0559 &  0.0040 &  49959 & cG & 159& Hsk& CANON X4 \\
RZ Cep              & 2458081.9253 & -0.0611 &  0.0015 &  50040 & cG & 153& Hsk& CANON X4 \\
RZ Cep              & 2458110.0175 & -0.0593 &  0.0023 &  50131 & cG & 170& Hsk& CANON X4 \\
DX Cep              & 2457611.2386 &  0.0241 &  0.0019 &  36033 & V  & 141& Hsk& 25cmL+ST402ME \\
EL Cep              & 2458056.9177 &  0.0524 &  0.0020 &  52751 & V  & 136& Hsk& 25cmL+ST402ME \\
EL Cep              & 2458078.9966 &  0.0501 &  0.0019 &  52804 & V  &  53& Hsk& 20cmL+ST402ME \\
EL Cep              & 2458081.9146 &  0.0517 &  0.0016 &  52811 & V  &  79& Hsk& 25cmL+ST402ME \\
EZ Cep              & 2457403.9207 &  0.1059 &  0.0010 &  81194 & V  & 188& Hsk& 25cmL+ST402ME \\
RV Cet              & 2457686.0295 &  0.2443 &  0.0040 &  29791 & V  & 160& Hsk& 13cmR+ST8XME \\
RZ Cet              & 2457997.2849 & -0.2218 &  0.0027 &  47180 & V  & 121& Hsk& 20cmL+ST402ME \\
HN Cet              & 2458079.0030 & -0.0211 &  0.0021 &  13285 & V  &  60& Hsk& 25cmL+ST402ME \\
HN Cet              & 2458109.9960 & -0.0272 &  0.0041 &  13352 & V  & 115& Hsk& 20cmL+ST402ME \\
S Com               & 2457463.2860 & -0.1115 &  0.0012 &  28655 & V  & 141& Hsk& 20cmL+ST402ME \\
V Com               & 2457500.9914 &  0.0528 &  0.0032 &  35846 & V  &  89& Hsk& 25cmL+ST402ME \\
RY Com              & 2457460.0486 &  0.2259 &  0.0017 &  38087 & V  & 244& Hsk& 20cmL+ST402ME \\
ST Com              & 2457867.0694 & -0.0440 &  0.0020 &  24448 & V  & 108& Hsk& 20cmL+ST402ME \\
TU Com              & 2457421.2036 & -0.1129 &  0.0022 &  61642 & V  & 179& Hsk& 25cmL+ST402ME \\
RV CrB              & 2457513.0815 & -0.1231 &  0.0016 &  43994 & V  & 159& Hsk& 13cmR+ST8XME \\
TV CrB              & 2457840.2905 &  0.0393 &  0.0017 &  44853 & V  &  94& Hsk& 20cmL+ST402ME \\
W Crt               & 2457500.9688 & -0.0325 &  0.0011 &  43340 & V  &  62& Hsk& 20cmL+ST402ME \\
W Crt               & 2457757.2428 & -0.0316 &  0.0010 &  43962 & V  & 126& Hsk& 13cmR+ST8XME \\
UY Cyg              & 2457513.1694 &  0.0692 &  0.0023 &  62563 & V  & 129& Hsk& 13cmR+ST8XME \\
UY Cyg              & 2458078.9248 &  0.0735 &  0.0018 &  63572 & V  & 111& Hsk& 20cmL+ST402ME \\
XZ Cyg              & 2457978.1435 &  0.1807 &  0.0016 &  29684 & cG & 137& Hsk& CANON X4 \\
DM Cyg              & 2457955.2492 &  0.0892 &  0.0025 &  36614 & V  &  86& Hsk& 20cmL+ST402ME \\
V759 Cyg            & 2457688.9847 & -0.0826 &  0.0020 &  56010 & V  & 145& Hsk& 20cmL+ST402ME \\
V759 Cyg            & 2458057.9959 & -0.1276 &  0.0012 &  57035 & V  & 107& Hsk& 20cmL+ST402ME \\
V1962 Cyg           & 2457997.1764 &  0.0150 &  0.0036 &   4461 & V  & 119& Hsk& 25cmL+ST402ME \\
ZZ Del              & 2457589.2194 &  0.0085 &  0.0020 &  38253 & V  & 179& Hsk& 20cmL+ST402ME \\
AX Del              & 2458026.9376 &  0.1774 &  0.0031 &  57159 & V  & 159& Hsk& 20cmL+ST402ME \\
BV Del              & 2458022.0818 &  0.0229 &  0.0022 &  76738 & V  & 146& Hsk& 20cmL+ST402ME \\
CK Del              & 2458057.9078 &  0.0881 &  0.0010 &  52981 & V  & 120& Hsk& 25cmL+ST402ME \\
DX Del              & 2458014.9824 &  0.0767 &  0.0032 &  39456 & cG & 161& Hsk& DCANON X4 \\
SU Dra              & 2457422.2325 &  0.0674 &  0.0023 &  20472 & cG & 161& Hsk& CANON KISS X3 \\
SU Dra              & 2457871.9782 &  0.0670 &  0.0024 &  21153 & cG & 186& Hsk& CANON X4 \\
SW Dra              & 2457449.9690 &  0.0635 &  0.0024 &  54813 & V  & 136& Hsk& 20cmL+ST402ME \\
XZ Dra              & 2457894.2219 & -0.1371 &  0.0015 &  33507 & cG & 129& Hsk& CANON X4 \\
BK Dra              & 2457863.2204 & -0.1682 &  0.0010 &  54621 & V  &  83& Hsk& 13cmR+ST8XME \\
BT Dra              & 2457824.2126 & -0.0225 &  0.0018 &  46018 & V  & 132& Hsk& 20cmL+ST402ME \\
RT Equ              & 2457953.2548 &  0.0873 &  0.0021 &  45150 & V  & 116& Hsk& 20cmL+ST402ME \\
RX Eri              & 2458021.2857 & -0.0062 &  0.0036 &  61863 & cG & 111& Hsk& CANON X4 \\
SV Eri              & 2457420.9897 &  0.3004 &  0.0041 &  30637 & V  & 100& Hsk& 13cmR+ST8XME \\
&  &  &  &  &   &   &  &  \\
\hline
\end{tabular}
\normalsize
\end{center}
\begin{center}
\small
\begin{tabular}{lrrrrlrll}\hline
star & max.HJD & O-C & err & E & band & n obs. & code & instrument\\
\hline
SV Eri              & 2457760.0625 &  0.3199 &  0.0045 &  31111 & V  & 125& Hsk& 13cmR+ST8XME \\
BN Eri              & 2457421.9408 &  0.0001 &  0.0014 &    131 & V  & 124& Hsk& 25cmL+ST402ME \\
BN Eri              & 2458109.0885 & -0.0001 &  0.0014 &   1540 & V  & 130& Hsk& 25cmL+ST402ME \\
LR Eri              & 2457420.9330 & -0.1479 &  0.0034 &   6663 & V  & 175& Hsk& 25cmL+ST402ME \\
SZ Gem              & 2457743.2225 & -0.0855 &  0.0011 &  60960 & V  & 108& Hsk& 25cmL+ST402ME \\
SZ Gem              & 2457758.2566 & -0.0855 &  0.0012 &  60990 & V  &  75& Hsk& 13cmR+ST8XME \\
V426 Gem            & 2458021.2847 & -0.0159 &  0.0025 &  12251 & V  & 119& Hsk& 20cmL+ST402ME \\
TW Her              & 2457463.2730 & -0.0163 &  0.0011 &  89884 & V  & 153& Hsk& 25cmL+ST402ME \\
VZ Her              & 2457422.3021 &  0.0795 &  0.0012 &  46859 & V  & 140& Hsk& 25cmL+ST402ME \\
VZ Her              & 2458014.9888 &  0.0849 &  0.0009 &  48205 & V  & 114& Hsk& 20cmL+ST402ME \\
AF Her              & 2457891.9784 & -0.1818 &  0.0031 &  47956 & V  & 174& Hsk& 20cmL+ST402ME \\
BD Her              & 2457513.1536 & -0.1604 &  0.0016 &  52511 & V  & 131& Hsk& 20cmL+ST402ME \\
BD Her              & 2458021.9639 &  0.1484 &  0.0032 &  53584 & V  & 171& Hsk& 25cmL+ST402ME \\
DL Her              & 2457490.2241 &  0.0523 &  0.0032 &  32591 & V  & 162& Hsk& 20cmL+ST402ME \\
DL Her              & 2458014.9952 &  0.0495 &  0.0012 &  33478 & V  & 106& Hsk& 25cmL+ST402ME \\
EE Her              & 2457513.1701 &  0.1287 &  0.0023 &  58744 & V  & 121& Hsk& 25cmL+ST402ME \\
EP Her              & 2457490.2545 & -0.1005 &  0.0020 &  69158 & V  & 157& Hsk& 25cmL+ST402ME \\
EP Her              & 2457542.1933 & -0.0997 &  0.0020 &  69280 & V  &  67& Hsk& 13cmR+ST8XME \\
EP Her              & 2458021.9691 & -0.1113 &  0.0023 &  70407 & V  & 170& Hsk& 20cmL+ST402ME \\
IP Her              & 2457921.2090 &  0.1920 &  0.0030 &  68876 & V  & 134& Hsk& 25cmL+ST402ME \\
V347 Her            & 2457978.1545 & -0.1522 &  0.0013 &  40710 & V  &  93& Hsk& 20cmL+ST402ME \\
V394 Her            & 2457513.2632 & -0.1894 &  0.0013 &  63675 & V  & 118& Hsk& 25cmL+ST402ME \\
V442 Her            & 2457508.2184 &  0.1685 &  0.0022 &  49261 & V  & 132& Hsk& 25cmL+ST402ME \\
V552 Her            & 2457508.2707 & -0.0005 &  0.0009 &    938 & V  &  96& Hsk& 25cmL+ST402ME \\
V1318 Her           & 2457846.2509 & -0.0329 &  0.0015 &   8170 & V  & 149& Hsk& 20cmL+ST402ME \\
SZ Hya              & 2457421.1619 & -0.2671 &  0.0025 &  31163 & V  & 136& Hsk& 13cmR+ST8XME \\
UU Hya              & 2457435.0820 &  0.0347 &  0.0023 &  34276 & V  & 137& Hsk& 20cmL+ST402ME \\
UU Hya              & 2457757.2549 &  0.0286 &  0.0032 &  34891 & V  & 114& Hsk& 20cmL+ST402ME \\
XX Hya              & 2457422.0628 & -0.0188 &  0.0014 &  34642 & V  & 207& Hsk& 20cmL+ST402ME \\
XX Hya              & 2457816.0766 & -0.0324 &  0.0014 &  35418 & V  & 141& Hsk& 20cmL+ST402ME \\
GO Hya              & 2457435.0053 & -0.0748 &  0.0079 &  49943 & V  & 219& Hsk& 13cmR+ST8XME \\
V496 Hya            & 2457460.1192 & -0.0189 &  0.0025 &   7490 & V  & 153& Hsk& 25cmL+ST402ME \\
CQ Lac              & 2457997.2860 &  0.1988 &  0.0011 &  37031 & V  &  91& Hsk& 25cmL+ST402ME \\
RR Leo              & 2457421.0821 &  0.1517 &  0.0012 &  31224 & V  & 123& Hsk& 13cmR+ST8XME \\
RR Leo              & 2457743.1933 &  0.1589 &  0.0010 &  31936 & V  & 125& Hsk& 13cmR+ST8XME \\
SS Leo              & 2457867.0734 & -0.1048 &  0.0014 &  25682 & V  & 115& Hsk& 13cmR+ST8XME \\
ST Leo              & 2457450.1023 & -0.0178 &  0.0013 &  61774 & V  & 154& Hsk& 13cmR+ST8XME \\
TV Leo              & 2457476.0500 &  0.1265 &  0.0018 &  30369 & V  & 160& Hsk& 25cmL+ST402ME \\
TV Leo              & 2457722.3154 &  0.1279 &  0.0018 &  30736 & V  & 162& Hsk& 20cmL+ST402ME \\
WW Leo              & 2457422.2242 &  0.0484 &  0.0029 &  37389 & V  & 176& Hsk& 20cmL+ST402ME \\
AA Leo              & 2457816.0360 & -0.1066 &  0.0014 &  30434 & V  & 147& Hsk& 25cmL+ST402ME \\
AE Leo              & 2457743.3117 &  0.1414 &  0.0022 &  60479 & V  & 118& Hsk& 20cmL+ST402ME \\
BO Leo              & 2457816.0542 &  0.0198 &  0.0022 &  35374 & V  & 144& Hsk& 13cmR+ST8XME \\
LL Leo              & 2457867.0934 & -0.0958 &  0.0020 &  18933 & V  & 105& Hsk& 25cmL+ST402ME \\
MR Leo              & 2457408.3514 & -0.0070 &  0.0017 &   6022 & V  &  63& Hsk& 20cmL+ST402ME \\
MR Leo              & 2457718.3422 &  0.0038 &  0.0021 &   6664 & V  & 107& Hsk& 20cmL+ST402ME \\
MR Leo              & 2457753.1002 & -0.0024 &  0.0025 &   6736 & V  & 181& Hsk& 25cmL+ST402ME \\
V LMi               & 2457434.9750 &  0.0276 &  0.0020 &  69688 & V  & 133& Hsk& 20cmL+ST402ME \\
X LMi               & 2457459.9758 &  0.2769 &  0.0026 &  26683 & V  & 106& Hsk& 20cmL+ST402ME \\
BF Lep              & 2457668.2663 & -0.0160 &  0.0020 &   7429 & V  & 135& Hsk& 25cmL+ST402ME \\
&  &  &  &  &   &   &  &  \\
\hline
\end{tabular}
\normalsize
\end{center}
\begin{center}
\small
\begin{tabular}{lrrrrlrll}\hline
star & max.HJD & O-C & err & E & band & n obs. & code & instrument\\
\hline
TV Lib              & 2457490.1813 & -0.0073 &  0.0008 & 138982 & V  & 114& Hsk& 13cmR+ST8XME \\
RW Lyn              & 2457420.9414 & -0.1966 &  0.0019 &  63060 & V  & 171& Hsk& 20cmL+ST402ME \\
RW Lyn              & 2457676.2041 & -0.2002 &  0.0021 &  63572 & V  & 135& Hsk& 20cmL+ST402ME \\
TV Lyn              & 2457676.2125 &  0.0328 &  0.0038 &  69500 & V  & 135& Hsk& 13cmR+ST8XME \\
TV Lyn              & 2458110.1068 &  0.0330 &  0.0028 &  71303 & V  & 114& Hsk& 25cmL+ST402ME \\
EP Lyn              & 2457671.1880 &  0.1815 &  0.0020 &  12889 & V  &  67& Hsk& 25cmL+ST402ME \\
RR Lyr              & 2457685.9265 &  0.1370 &  0.0022 &  26042 & cG & 156& Hsk& CANON X3 \\
RR Lyr              & 2457955.1464 &  0.0947 &  0.0035 &  26517 & cG & 130& Hsk& CANON X4 \\
RR Lyr              & 2458022.0261 &  0.0840 &  0.0027 &  26635 & cG & 187& Hsk& CANON X4 \\
RR Lyr              & 2458063.9665 &  0.0762 &  0.0017 &  26709 & cG & 170& Hsk& CANON X4 \\
CN Lyr              & 2458015.0770 &  0.0242 &  0.0040 &  32886 & V  &  64& Hsk& 13cmR+ST8XME \\
FN Lyr              & 2457513.2593 &  0.0330 &  0.0025 &  44907 & V  & 112& Hsk& 20cmL+ST402ME \\
KX Lyr              & 2457685.9195 &  0.0123 &  0.0020 &  40950 & V  & 160& Hsk& 20cmL+ST402ME \\
V895 Mon            & 2457679.2953 & -0.0424 &  0.0025 &   9234 & V  & 137& Hsk& 20cmL+ST402ME \\
ST Oph              & 2457508.1646 & -0.0267 &  0.0012 &  64575 & V  & 120& Hsk& 20cmL+ST402ME \\
ST Oph              & 2457902.2272 & -0.0258 &  0.0019 &  65450 & V  & 195& Hsk& 20cmL+ST402ME \\
V445 Oph            & 2457823.3265 &  0.0492 &  0.0010 &  76267 & V  & 119& Hsk& 20cmL+ST402ME \\
V452 Oph            & 2457863.2733 &  0.0012 &  0.0016 &  38060 & V  &  90& Hsk& 20cmL+ST402ME \\
V455 Oph            & 2457824.3312 &  0.1257 &  0.0022 &  35229 & V  & 111& Hsk& 20cmL+ST402ME \\
V816 Oph            & 2457863.2057 & -0.1484 &  0.0017 &  56507 & V  &  67& Hsk& 20cmL+ST402ME \\
CM Ori              & 2457760.0431 &  0.0098 &  0.0020 &  49490 & V  & 144& Hsk& 20cmL+ST402ME \\
CM Ori              & 2458109.0011 &  0.0173 &  0.0026 &  50022 & V  & 125& Hsk& 20cmL+ST402ME \\
V964 Ori            & 2457780.0725 & -0.0319 &  0.0017 &  52075 & V  &  85& Hsk& 25cmL+ST402ME \\
V964 Ori            & 2458109.0936 & -0.0466 &  0.0011 &  52727 & V  & 126& Hsk& 20cmL+ST402ME \\
VV Peg              & 2457588.2412 & -0.0013 &  0.0013 &  37263 & V  & 130& Hsk& 20cmL+ST402ME \\
VV Peg              & 2458063.9363 &  0.0048 &  0.0014 &  38237 & V  & 116& Hsk& 20cmL+ST402ME \\
AV Peg              & 2457542.2184 &  0.1728 &  0.0011 &  35227 & V  & 153& Hsk& 20cmL+ST402ME \\
AV Peg              & 2458027.0743 &  0.1834 &  0.0009 &  36469 & V  & 127& Hsk& 13cmR+ST8XME \\
BF Peg              & 2457609.2949 & -0.0735 &  0.0031 &  29543 & V  & 188& Hsk& 20cmL+ST402ME \\
BF Peg              & 2458052.0397 & -0.0937 &  0.0018 &  30436 & V  & 149& Hsk& 20cmL+ST402ME \\
BH Peg              & 2458062.9000 & -0.1458 &  0.0024 &  29170 & V  & 134& Hsk& 20cmL+ST402ME \\
CG Peg              & 2457974.2440 & -0.0681 &  0.0019 &  40399 & V  & 136& Hsk& 20cmL+ST402ME \\
CV Peg              & 2458051.9260 & -0.0722 &  0.0021 &  59173 & V  & 125& Hsk& 25cmL+ST402ME \\
CY Peg              & 2458022.1176 &  0.3216 &  0.0017 &  51319 & V  & 136& Hsk& 13cmR+ST8XME \\
DZ Peg              & 2458082.0077 &  0.1842 &  0.0020 &  39830 & V  & 112& Hsk& 20cmL+ST402ME \\
ES Peg              & 2457921.2062 &  0.1881 &  0.0013 &  37165 & V  & 159& Hsk& 20cmL+ST402ME \\
ES Peg              & 2458104.8964 &  0.1912 &  0.0023 &  37506 & V  &  53& Hsk& 20cmL+ST402ME \\
ET Peg              & 2457542.2033 & -0.0679 &  0.0014 &  37842 & V  & 167& Hsk& 25cmL+ST402ME \\
ET Peg              & 2457747.9375 & -0.0640 &  0.0023 &  38262 & V  & 105& Hsk& 20cmL+ST402ME \\
ET Peg              & 2458057.0204 & -0.0664 &  0.0025 &  38893 & V  &  73& Hsk& 25cmL+ST402ME \\
IY Peg              & 2457974.1741 &  0.0054 &  0.0021 &  13333 & V  & 109& Hsk& 25cmL+ST402ME \\
V453 Peg            & 2457686.0540 & -0.0133 &  0.0018 &   6442 & V  & 185& Hsk& 20cmL+ST402ME \\
V453 Peg            & 2458027.0820 & -0.0082 &  0.0028 &   7022 & V  & 155& Hsk& 25cmL+ST402ME \\
V509 Peg            & 2457997.1740 &  0.0153 &  0.0010 &  12954 & V  & 120& Hsk& 20cmL+ST402ME \\
V509 Peg            & 2458109.9200 &  0.0149 &  0.0011 &  13263 & V  &  98& Hsk& 25cmL+ST402ME \\
V606 Peg            & 2458064.0069 & -0.0099 &  0.0018 &   7088 & V  & 115& Hsk& 20cmL+ST402ME \\
TU Per              & 2457668.1735 & -0.2298 &  0.0018 &  31126 & V  &  87& Hsk& 13cmR+ST8XME \\
AR Per              & 2458090.9860 &  0.0692 &  0.0020 &  72504 & V  & 102& Hsk& 20cmL+ST402ME \\
ET Per              & 2457759.9342 &  0.0448 &  0.0014 &  75065 & V  & 143& Hsk& 20cmL+ST402ME \\
ET Per              & 2458057.0174 &  0.0417 &  0.0020 &  75819 & V  &  88& Hsk& 20cmL+ST402ME \\
&  &  &  &  &   &   &  &  \\
\hline
\end{tabular}
\normalsize
\end{center}
\begin{center}
\small
\begin{tabular}{lrrrrlrll}\hline
star & max.HJD & O-C & err & E & band & n obs. & code & instrument\\
\hline
FM Per              & 2457753.1061 & -0.1943 &  0.0012 &  49034 & V  & 175& Hsk& 20cmL+ST402ME \\
FM Per              & 2458104.9102 & -0.1261 &  0.0021 &  49753 & V  &  47& Hsk& 25cmL+ST402ME \\
V375 Per            & 2457759.9183 &  0.1686 &  0.0024 &  53510 & V  & 143& Hsk& 25cmL+ST402ME \\
FF Psc              & 2458109.9943 &  0.0431 &  0.0017 &   9480 & V  & 137& Hsk& 25cmL+ST402ME \\
FR Psc              & 2457753.0244 &  0.0265 &  0.0022 &   8869 & V  &  85& Hsk& 25cmL+ST402ME \\
FR Psc              & 2458108.9106 &  0.0266 &  0.0014 &   9650 & V  & 130& Hsk& 20cmL+ST402ME \\
HT Psc              & 2457752.9165 & -0.0503 &  0.0012 &   7317 & V  & 110& Hsk& 25cmL+ST402ME \\
HT Psc              & 2458052.0244 & -0.0510 &  0.0017 &   7862 & V  & 152& Hsk& 25cmL+ST402ME \\
HT Psc              & 2458108.9994 & -0.0491 &  0.0021 &   7967 & V  & 110& Hsk& 25cmL+ST402ME \\
XX Pup              & 2457404.0653 &  0.0489 &  0.0015 &  30223 & V  & 153& Hsk& 13cmR+ST8XME \\
BB Pup              & 2457815.9609 &  0.1502 &  0.0019 &  39549 & V  & 123& Hsk& 20cmL+ST402ME \\
VY Ser              & 2457863.2474 &  0.0639 &  0.0048 &  37303 & cG & 120& Hsk& CANON X4 \\
AN Ser              & 2457513.0813 &  0.0012 &  0.0012 &  81989 & V  & 120& Hsk& 25cmL+ST402ME \\
AV Ser              & 2457490.1818 &  0.1783 &  0.0019 &  59781 & V  & 100& Hsk& 25cmL+ST402ME \\
AV Ser              & 2457902.1727 &  0.1832 &  0.0015 &  60626 & V  & 100& Hsk& 13cmR+ST8XME \\
AW Ser              & 2457508.2766 &  0.0458 &  0.0016 &  48843 & V  & 123& Hsk& 20cmL+ST402ME \\
BH Ser              & 2457463.2461 &  0.1436 &  0.0011 &  36775 & V  & 110& Hsk& 13cmR+ST8XME \\
CS Ser              & 2457513.0876 &  0.0234 &  0.0028 &  49994 & V  & 121& Hsk& 20cmL+ST402ME \\
DF Ser              & 2457496.0768 &  0.1020 &  0.0020 &  63476 & V  &  66& Hsk& 20cmL+ST402ME \\
V423 Ser            & 2457463.1896 & -0.0145 &  0.0021 &   7988 & V  & 124& Hsk& 20cmL+ST402ME \\
V Sex               & 2457725.3494 &  0.0420 &  0.0031 &  61908 & V  &  91& Hsk& 20cmL+ST402ME \\
SS Tau              & 2457641.2410 &  0.1580 &  0.0018 &  50303 & V  & 141& Hsk& 25cmL+ST402ME \\
SS Tau              & 2458091.0587 &  0.1761 &  0.0017 &  51519 & V  & 125& Hsk& 20cmL+ST402ME \\
IY Tau              & 2457449.9541 &  0.1664 &  0.0028 &  83520 & V  & 137& Hsk& 25cmL+ST402ME \\
IY Tau              & 2458082.0953 &  0.1813 &  0.0042 &  85199 & V  & 110& Hsk& 25cmL+ST402ME \\
U Tri               & 2457671.2701 & -0.0561 &  0.0015 &  86240 & V  & 106& Hsk& 20cmL+ST402ME \\
U Tri               & 2457779.9502 & -0.0585 &  0.0014 &  86483 & V  & 123& Hsk& 20cmL+ST402ME \\
RV UMa              & 2457421.1932 &  0.1316 &  0.0014 &  26376 & cG & 196& Hsk& CANON KISS X3 \\
TU UMa              & 2457444.3203 & -0.0630 &  0.0017 &  26204 & V  &  96& Hsk& 13cmR+ST8XME \\
TU UMa              & 2457743.2251 & -0.0632 &  0.0018 &  26740 & V  & 126& Hsk& 0cmL+ST402ME \\
AB UMa              & 2457422.2141 &  0.1267 &  0.0047 &  35349 & V  & 152& Hsk& 25cmL+ST402ME \\
EX UMa              & 2457460.0215 & -0.0064 &  0.0027 &   4111 & V  & 193& Hsk& 13cmR+ST8XME \\
ST Vir              & 2457824.2138 & -0.1400 &  0.0011 &  41594 & V  & 131& Hsk& 13cmR+ST8XME \\
UU Vir              & 2457872.0214 &  0.0112 &  0.0011 &  33798 & V  & 159& Hsk& 20cmL+ST402ME \\
UV Vir              & 2457476.0557 &  0.0218 &  0.0033 &  29779 & V  & 150& Hsk& 13cmR+ST8XME \\
UZ Vir              & 2457463.1544 &  0.1308 &  0.0012 &  70740 & V  & 140& Hsk& 25cmL+ST402ME \\
AE Vir              & 2457422.3188 &  0.1232 &  0.0029 &  45972 & V  & 119& Hsk& 20cmL+ST402ME \\
AF Vir              & 2457892.0018 &  0.1616 &  0.0015 &  36295 & V  & 183& Hsk& 25cmL+ST402ME \\
AT Vir              & 2457490.1660 &  0.1369 &  0.0020 &  33876 & V  & 103& Hsk& 20cmL+ST402ME \\
AV Vir              & 2457872.0120 &  0.0251 &  0.0029 &  24968 & V  & 162& Hsk& 25cmL+ST402ME \\
BC Vir              & 2457866.9848 &  0.2771 &  0.0015 &  67216 & V  & 138& Hsk& 25cmL+ST402ME \\
V388 Vir            & 2457421.2787 &  0.0203 &  0.0034 &   6515 & V  & 117& Hsk& 13cmR+ST8XME \\
V419 Vir            & 2457758.3512 & -0.0798 &  0.0036 &   9852 & V  & 103& Hsk& 20cmL+ST402ME \\
V476 Vir            & 2457512.9810 &  0.0614 &  0.0023 &   4335 & V  & 103& Hsk& 20cmL+ST402ME \\
CE Vul              & 2458026.9511 & -0.1135 &  0.0015 &  56279 & V  & 156& Hsk& 25cmL+ST402ME \\
FH Vul              & 2457955.1570 & -0.1674 &  0.0030 &  54087 & V  &  88& Hsk& 20cmL+ST402ME \\
FH Vul              & 2458056.9147 & -0.1698 &  0.0020 &  54338 & V  & 144& Hsk& 20cmL+ST402ME \\
ASAS J020058+1332.8 & 2457697.1929 &  0.0019 &  0.0033 &      0 & V  &  79& Hsk& 25cmL+ST402ME \\
ASAS J020058+1332.8 & 2457760.0394 & -0.0005 &  0.0024 &    102 & V  &  80& Hsk& 25cmL+ST402ME \\
ASAS J020058+1332.8 & 2458090.9177 & -0.0034 &  0.0025 &    639 & V  &  95& Hsk& 25cmL+ST402ME \\
&  &  &  &  &   &   &  &  \\
\hline
\end{tabular}
\normalsize
\end{center}
\begin{center}
\small
\begin{tabular}{lrrrrlrll}\hline
star & max.HJD & O-C & err & E & band & n obs. & code & instrument\\
\hline
NSV 1443            & 2458110.0794 &  0.0030 &  0.0019 &    770 & V  & 115& Hsk& 13cmR+ST8XME \\
NSV 1443            & 2458082.0705 & -0.0006 &  0.0017 &    724 & V  & 105& Hsk& 13cmR+ST8XME \\
NSV 1443            & 2457641.2922 &  0.0000 &  0.0030 &      0 & V  & 102& Hsk& 20cmL+ST402ME \\
NSV 3833            & 2458060.2970 & -0.0008 &  0.0030 &    745 & V  &  80& Hsk& 20cmL+ST402ME \\
NSV 3833            & 2457699.2619 &  0.0089 &  0.0040 &      4 & V  & 172& Hsk& 20cmL+ST402ME \\
NSV 3833            & 2457717.2669 & -0.0139 &  0.0015 &     41 & V  & 208& Hsk& 25cmL+ST402ME \\
NSV 3833            & 2457422.0131 & -0.0003 &  0.0025 &   -565 & V  & 259& Hsk& 13cmR+ST8XME \\
NSV 3833            & 2457697.3064 &  0.0024 &  0.0037 &      0 & V  & 158& Hsk& 25cmL+ST402ME \\
&  &  &  &  &   &   &  &  \\
\hline
\end{tabular}
\normalsize
\end{center}

\vspace{1.2cm}

\rule[.04in]{6in}{.02in} \\
VSOLJ \\
c/o Keiichi Saijo National Science Museum, Ueno-Park, Tokyo Japan \\ \\
Editor Seiichiro Kiyota \\
e-mail: skiyotax@gmail.com \\
Publishing Masahiko Momose \\ \\ \\
GEOS (Groupe Europ\'een d'Observations Stellaires) \\
23 Parc de Levesville, 28300 Bailleau l'Ev\^eque, France \\ \\
Editor J.F. Le Borgne\\
e-mail: jean-francois.leborgne@irap.omp.eu \\
\rule[.04in]{6in}{.02in}
 \\


\vspace{0.6cm}

\begin{thebibliography}{}

\bibitem[Klotz et al., 2009]{Klotz}
Klotz A., Bo\"er M., Atteia J.~L., Gendre B.,
Astronomical Journal, 2009, 137, 4100

\bibitem[Le Borgne et al., 2007]{Le Borgne 2007}
Le Borgne, J. F., et al., Astronomy and Astrophysics 476, 307
(2007)

\bibitem[Le Borgne et al., 2014]{Le Borgne 2014}
Le Borgne J.F., Poretti E., Klotz A., Denoux E., Smith H.A., Kolenberg K., Szab{\'o} R.,
Bryson S., Audejean M., Buil C., Caron J., Conseil E., Corp L., Drillaud C., de France T.,
Graham K., Hirosawa K., Klotz A.~N., Kugel F., Loughney D., Menzies K., Rodr{\'{\i}}guez M.,
Ruscitti P.~M.,
Monthly Notices of the Royal Astronomical Society, 2014, 441, 1435

\bibitem[Poretti et al., 2018]{Poretti}
Poretti E., Le Borgne J.F., Klotz A., Rainer M., Correa M., 2018, eprint arXiv:1801.09702,
Contributed talk at the "RR Lyrae 2017 Conference - Revival of the Classical Pulsators:
from Galactic Structure to Stellar Interior Diagnostics" (Niepolomice, Poland, 17-21 September, 2017)

\bibitem[Preston et al, 1965]{Preston}
Preston G.~W., Smak J., Paczynski B.,
Astrophysical Journal Supplement, 1965, 12, 99

\bibitem[Samus et al., 2017]{GCVS}
Samus N.N., Kazarovets E. V., Durlevich O.V., Kireeva N.N., Pastukhova E.N.,
General Catalogue of Variable Stars: new version. GCVS 5.1
(the first stage of the fifth edition), ARep,2017,60, №1

\bibitem[Vandenbroere et al., 2014]{Vandenbroere}
Vandenbroere J., Le Borgne J.-F., Boninsegna R., 2014, GEOS Circular RR53
\end{thebibliography}
\end{document}